\documentclass[%
 aip,
 amsmath,amssymb,
 reprint,%
]{revtex4-1}

\usepackage{graphicx}
\usepackage{dcolumn}
\usepackage{bm}

\usepackage[utf8]{inputenc}
\usepackage[T1]{fontenc}
\usepackage{mathptmx}

\usepackage{graphicx,color}
\usepackage{dcolumn}
\usepackage{bm}
\usepackage{psfrag}
\usepackage{amsmath,amssymb}
\usepackage{nccfoots}
\usepackage{stmaryrd}
\usepackage{float}
\usepackage[normalem]{ulem}
\usepackage{xcolor}

\newcommand{\green}[1]{{\textcolor{black}{#1}}}
\newcommand{\blue}[1]{{\textcolor{black}{#1}}}
\newcommand{\Nblue}[1]{{\textcolor{black}{#1}}}
\newcommand\redsout{\bgroup\markoverwith{\textcolor{red}{\rule[0.5ex]{2pt}{0.4pt}}}\ULon}
\newcommand\bluesout{\bgroup\markoverwith{\textcolor{blue}{\rule[0.5ex]{2pt}{0.4pt}}}\ULon}

\begin{document}

\preprint{/ \textit{to the Journal of Chemical Physics, Special Topic on Quantum Light}}

\title{Light-matter quantum dynamics of complex laser-driven systems}

\author{Ivan~Gonoskov}
\email{ivan.gonoskov@uni-jena.de} 
\author{Stefanie~Gräfe}
\email{s.graefe@uni-jena.de}
\affiliation{Institute for Physical Chemistry, Friedrich-Schiller-University Jena, Max-Wien-Platz 1, 07743 Jena, Germany}\date{13 June 2021}

\begin{abstract}
We propose a novel general approximation to transform and simplify the description of a complex fully-quantized system describing the interacting light and matter. The method has some similarities to the time-dependent Born-Oppenheimer \Nblue{approach}: we consider a quantum description of light rather than of nuclei and follow a similar separation procedure. Our approximation allows to obtaing a decoupled system for the light-excited matter and ''dressed'' light connected parametrically. With these equations at hand, we study how intense light as a quantum state is affected due to the back-action of the interacting matter. We discuss and demonstrate the possibility of the light-mode entanglement and nonclassical light generation during the interaction.
\end{abstract}

\maketitle

{\textit{Motivation.}-- Light is a unique tool to control and probe the world of matter/particles because of its fascinating properties: nowadays, technologies allow to generate, measure, and control light radiation at extremely precise level, both on spatial and temporal scales. This opens extraordinary possibilities to explore the quantum nature of our world at the micro/nanometer spatial scales, attosecond time scales, and possibly will bring some new quantum phenomena to macroscopic level. The quantum-mechanical description of both, the matter (charged particles) and the electromagnetic radiation is universal: within conventional non-relativistic quantum theory, everything can be described by the total wave function which is a solution of the corresponding time-dependent Scr\"odinger equation (TDSE) with a complete Hamiltonian including all the interactions. Since light is a quantum object \cite{Compton}, the complete and consistent treatment requires a fully-quantum description for the whole interacting system including both, the charged particles and the light modes. It is known that the calculation of the complex quantum problems is a very complicated issue, especially for strongly interacting, multidimensional systems. Here, we present a novel theoretical method which simplifies the problem significantly, while still taking various purely quantum properties into account.}

{One of the most general and rigorous approaches which combines classical electrodynamics (non-relativistic for the particles) and quantum principles within the time-dependent Schr\"odinger equation (TDSE) employs the so-called \textit{Pauli-Fierz Hamiltonian} (PFH), see \cite{PFH} and equation (1), below. This Pauli-Fierz Hamiltonian contains the description of the dynamics of the charged particles coupled with the electromagnetic fields in a fully quantum way. This unifies Maxwell's equations and Newton's equations at the fully quantum description. Within this theory, it is possible also to include the spin of the particles and several semi-relativistic effects \cite{PFH2,PFH3}. The possible solutions of this TDSE could open various opportunities for applications ranging from new sources of light or states of quantum matter to unique pure quantum collective effects. Recent experiments demonstrate some surprising quantum collective phenomena beyond classical \newline or even semiclassical pictures \cite{wP0,quantExp}.}

The paper is organized as follows: we start with the TDSE description with the PFH. Next, we derive the related simplified system of the parametrically-connected equations. Finally, we consider selected examples/solutions and discuss the consequences and the possible applications.


\textit{General description of the light-matter interaction.}-- The total Hamiltonian for the charged particles (numbered with indices $k$ and $k'$ ) and the multi-mode electromagnetic field ({modes} numbered with index $j$) in the Schr\"odinger representation \Nblue{with Coulomb gauge} reads as follows {(known as Pauli-Fierz Hamiltonian, see overview and details in \cite{PFH})}:
\begin{equation}\label{pf12H}
\hat{H}=\sum\limits_{k}\frac{1}{2m_{k}}\left(\,\hat{\vec{p}}_{k}-\frac{e_{k}}{c}\sum\limits_{j}\hat{\vec{A}}_{j}\right)^{2}+\sum\limits_{k\,k'}U_{k\,k'}+\sum\limits_{j}\omega_{j}\hat{N}_{j}\;.
\end{equation}
Here, $e_{k}$ and $m_{k}$ are the charges and masses of the particles with index $k$\,, $\sum\limits_{k\,k'}U_{k\,k'}$ includes particle-particle mutual interactions, as well as interaction of each particle with external {longitudinal} fields. {Each} mode $j$ of the electromagnetic field is quantized {via the term} $\omega_{j}\hat{N}_{j}$ in the Hamiltonian. This is the most general formulation, {without including spin (which can be extended separately within Pauli-Fierz Hamiltonian) within a non-relativistic description of matter (particles). 

\blue{\textit{The idea of parametrical-connection approximation:} Based on the PFH, we first consider the quantum interaction of two objects of different nature (one electron and the light field) having the total Hamiltonian of the form: $\hat{H}=\hat{H}_{e}+\hat{W}_{int}+\hat{H}_{field}$.  We also consider that the initial state of the whole system is of the form: $\Psi_{0}(x,q)=F_{0}(x)\otimes{}G_{0}(q)$, where $x$ is related to the electronic coordinates and $q$ is related to the field-modes coordinates, respectively. In the strong-field case, when $\left\langle \hat{H}_{e} \right\rangle \sim \left\langle \hat{W}_{int} \right\rangle \ll \left\langle \hat{H}_{field} \right\rangle$, it is reasonable to assume that the quantum state of the light field changes relatively weakly during the interaction compare to the initial state $G_{0}(q)$ since the interacting electron is a small perturbation to the strong field. This leads to the idea of the following (full first-order) approximation of the total solution: $\Psi(x,q,t)\approx{}F_{(G_{0})}(x,q,t)\otimes{}G_{(F)}(q,t)$, with the error proportional to some degree of the ratio $\left\langle \hat{W}_{int} \right\rangle / \left\langle \hat{H}_{field} \right\rangle$. The most signficant simplification in the following is that the solution for the electronic subsystem $F_{(G_{0})}(x,q,t)$ is obtained from the TDSE with only local operators on $q$ (there are no derivatives), i.e. it is an effectively one-object TDSE with parametrical $q$-dependance. The corresponding $q$-derivatives can be transformed to the simple $q$-dependent expressions using a non-perturbed $G_{0}(q)$ within this first-order approximation. Thus, we transform the initial two-object TDSE to a simplified system of two equations, one for the $F_{(G_{0})}(x,q,t)$, where $q$ is a parameter, and other for the field-state evolution $G_{(F)}(q,t)$ including a small electron back-action via $F$. The detailed derivation of the system for coherent (Gaussian) initial state of the intense light is given below. The weaker approximation: $\Psi(x,q,t)\approx{}F(x,\{ q\},t)\otimes{}G_{0}$ leads immediately after the field-state averaging of the total Hamiltonian in the high photon number limit to the known (fixed-amplitude) classical field limit \cite{ClassFL}.}

\blue{\textit{Detailed derivation of the general approximation:}} {We first} consider one particle - electron interacting with \Nblue{an arbitrary given configuration of propagating electromagnetic} plane waves. \Nblue{A similar problem statement was recently considered for a cavity-photon Hamiltonian in the dipole approximation\cite{hoffm}. We note, our approach can be transformed for more complex systems with many particles, as well as for more specific ones.} {(atomic units $\hbar=m_{e}=|e|=1$ are used in what follows).} We {use the} coordinate representation of the field operators \cite{quant1,quant2,quant3}, with the creation and annihilation operators $\hat{a}^{+}_{j}=\frac{1}{\sqrt{2}}\left(q_{j}-\frac{\partial}{\partial{q_{j}}}\right)$ and $\hat{a}_{j}=\frac{1}{\sqrt{2}}\left(q_{j}+\frac{\partial}{\partial{q_{j}}}\right)$ and the field dynamical coordinate $q_{j}$. {We assume that the initial state \green{(at the time $t=0$)} of each mode \green{$j$} is coherent: $G_{j}(\green{q_{j},\,}0)=\green{G_{0\,j}=}\;C_{0\,j}\cdot\exp\left[-\frac{1}{2}(q_{j}-e^{-i\theta_{j}}\sqrt{2N_{j}})^{2}\right]$ with $N_{j}$ being an initial arbitrary average number of photons, and $\theta_{j}$ is initial phase (see Appendix). In this case, if we apply the following transformation $\prod\limits_{j}\exp[-i(\omega_{j}t+\theta_{j})\hat{N}_{j}]$ (see also \cite{quant1}) to the general TDSE with Hamiltonian Eq.(\ref{pf12H}), we obtain the TDSE in the interaction picture with phase-dependent field operators and phase-independent mode states: 
\blue{
\begin{equation}\label{Hn}
i\dot{\Psi}=\hat{H}\Psi\;,\;\;\;\;\hat{H}=\frac{1}{2}\left[\,\hat{\vec{p}}-\frac{1}{c}\sum\limits_{j}\hat{\vec{A}}_{j}\right]^{2}+U(\vec{r},t)\;.
\end{equation}
}
{The initial state \green{for each mode $j$} \blue{of the light field} in this representation is $G_{j}(\green{q_{j},\,}0)=C_{0\,j}\cdot\exp\left[-\frac{1}{2}(q_{j}-\sqrt{2N_{j}})^{2}\right]$, \Nblue{the TDSE for the field states in vacuum is $i\dot{G}_{j}^{\;vac}=0$}, and the quantized field vector potential operator reads: 
\begin{equation}\label{pfA}
\begin{aligned}
\hat{\vec{A}}_{j}=&{\vec{\chi}}_{j}\;\beta_{j}q_{j}\;\cos(\vec{\kappa}_{j}\vec{r}-\omega_{j}t-\theta_{j})\\
&-i{\vec{\chi}}_{j}\;\beta_{j}\frac{\partial}{\partial{}q_{j}}\;\sin(\vec{\kappa}_{j}\vec{r}-\omega_{j}t-\theta_{j})\,,
\end{aligned}
\end{equation}
\Nblue{where ${\vec{\chi}}_{j}$ are polarization vectros (linear in this case), and $\beta_{j}=c\sqrt{2\pi/\omega_{j}{V}}$ are parameters (with the same dimensionality as ${A}_{j}$) which depend on quantization volume $V$, according to the conventional plane-wave quantization procedure \cite{quant1}.} In our case, when the initial mode state is a coherent state, \Nblue{each $\beta_{j}$} parameter naturally connects the field mode amplitude $A_{0\,j}$ and the initial average photon number in this mode $N_{j}$ as follows $A_{0\,j}=\beta_{j}\sqrt{2N_{j}}$, \blue{and consequently, $\beta\sim\left[\left\langle \hat{W}_{int} \right\rangle / \left\langle \hat{H}_{field} \right\rangle \right]^\frac{1}{2}$}. \Nblue{A typical laser pulse, which is finite in space and time and consists of a number of modes, can be described in terms of an effective quantization volume $V_{eff}$. This volume is related to the sizes of the radiation focusing area and the effective interaction volume, see also \cite{quant1,nqvol}.} \Nblue{Based on the estimations for the number of relevant experiments with strong laser pulses interacting with gases or solids, where quantum-optical effects are measurable \cite{wP0,quantExp}, the effective values are $(\beta_{j}/A_{0j})\sim{}10^{-7}$ or even much less.} These parameters are also responsible for the quantum dispersion of the quantized radiation in the presence of charged particles, see \cite{my1q}. We refer $\beta_{j}$ as the fundamental scaling parameters for the corresponding light-matter interaction problem, \Nblue{when $E_{0j}\ll{}E_{atomic}$ in the optical wavelength range}. Due to its very small values, we may conclude from Eq.(\ref{Hn}) that even in a strong-field case, the back-action of one electron to the light-mode wavefunction is small and depends somehow on $\beta_{j}$. \Nblue{Therefore, the state of electromagnetic field $G(q_{j},t)$ is affected as $G\approx{}G_{0}+\sum{}O(\beta_{j})$. Finding the equation for $G$ is a key point of our study.} 


It was rigorously demonstrated that the solution of the two-component TDSE \Nblue{with an arbitrary Hamiltonian} can be written as a product two special wavefunctions each satisfying the normalization condition, see \cite{EF1} for full details. \Nblue{In our case, it leads to the following exact factorization: $\Psi=F_{ex}(\vec{r},q_{j},t)\cdot{}G_{ex}({q_{j},t})$, with the related system of equations for the $F_{ex}$ and $G_{ex}$ of the same complexity as the original TDSE, see also \cite{EF1}. Instead, based on our previous considerations regarding the dependance on parameters $\beta_{j}$, we consider the solution of Eq.(\ref{Hn}) of the form $\Psi\approx{}F_{G_{0}}(\vec{r},q_{j},\beta_{j},t)\cdot{}G({q_{j},\beta_{j},t})+ \sum{}O(\beta_{j}^{s})$.} In our approximation, we have the same form of the solution with the same partial normalization conditions: $\left\langle F|F \right\rangle = \int{}|F(\vec{r},q_{j},t)|^{2}d^{3}r\equiv{}1$, $\left\langle G|G \right\rangle=\int{}|G(q_{j},t)|^{2}[d\,q_{j}]\equiv{}1$. The representation is not unique since one can construct others one by choosing $\Psi=F(\vec{r},q_{j},t)e^{i\chi(q_{j},t)}\,\cdot\,G(q_{j},t)e^{-i\chi(q_{j},t)}$ with an arbitrary function $\chi(q_{j},t)$ matching the initial condition at $t=0$, see also \cite{EF1}. In order to have a uniqueness of the representation for a given initial state, we require the condition: $\left\langle F|\dot{F} \right\rangle =0$, which is often referred to as a static gauge of the factorization \cite{ClassFL}.

\blue{Using the considered form of the solution, we can write:
\begin{equation}\label{Hn2}
i\dot{\Psi}=i\dot{F}G+iF\dot{G}=\hat{H}\,FG\;,
\end{equation}
and obtain the decoupling in first-order approximation:
\begin{subequations}\label{Hn3}
\begin{align}
i\dot{F}=&\left[G^{-1}\hat{H}G\right]\,F-i\,\frac{\dot{G}}{G}\,F\approx\left[G_{0}^{-1}\hat{H}G_{0}\right]\,F-i\phi\,F\,,\\
&\;\text{since:}\;G=G_{0}+\sum_{j}O(\beta_{j})\;,\;t>0\,;\nonumber\\
i\dot{G}=&\left\langle F\left|\hat{H}\right|F \right\rangle\,G\,,\;\text{under the static gauge:}\,\left\langle F|\dot{F} \right\rangle =0\,.
\end{align}
\end{subequations}
}From this it follows that in a first step of a reasonable approximation, for the electron subsystem, we may use $\beta_{j}$-independent states of light modes for the approximation of the vector potential operator. \blue{Thus, we can exclude the non-local operator $\frac{\partial}{\partial{}q_{j}}$ in the equation for the electron subsystem \Nblue{(since $\beta_{j}\frac{\partial}{\partial{}q_{j}}F\sim{}\beta_{j}^{2}$)}, having instead a simple local operator as follows: $\beta_{j}\frac{\partial}{\partial{}q_{j}}\rightarrow\{\beta_{j}G_{0\,j}^{-1}\frac{\partial}{\partial{}q_{j}}G_{0\,j}\}$}. \green{Starting from Eq.(\ref{Hn3}a), we assume that $G(q,t)$ does not have zeros within a time interval of consideration. This can be in the case when one starts with coherent state and considers conditions of relatively small deviations, leading for example to the various nonclassical/squeezed states. Otherwise the procedure can be modified in a way that instead of Eq.(\ref{Hn3}a) we can write $iG\dot{F}=\hat{H}\;G_{0}F - i\dot{G}F$ with the similar subsequent related considerations.} The described approach allows us to decouple the total system Eq.(\ref{Hn}) and separate the equations for the electronic subsystem and ''dressed'' light subsystem via $\beta_{j}$ parametrical connections. We obtain finally for the quantum light-electron interaction problem:}\\
\begin{subequations}\label{pf13H}
\begin{align}
&i\dot{F}=\hat{H}_{P}F\;,\;\;\;\;\hat{H}_{P}=\frac{1}{2}\left[\,\hat{\vec{p}}-\frac{1}{c}\sum\limits_{j}\vec{A}_{j}^{par}\right]^{2}+U(\vec{r},t)\;,\\
&i\dot{G}=\hat{H}_{B}G\;,\;\;\;\;\Nblue{\hat{H}_{B}=\left\langle F\left|\hat{H}\right|F \right\rangle}\;,
\end{align}
\end{subequations}
{where $\vec{A}_{j}^{par}=\green{\vec{\bf \chi}_{j}}\{\beta_{j}q_{j}\}\cos(\vec{\kappa}_{j}\vec{r}-\omega_{j}t-\theta_{j})-i\green{\vec{\bf \chi}_{j}}\{\beta_{j}G_{0\,j}^{-1}\frac{\partial}{\partial{}q_{j}}G_{0\,j}\}\sin(\vec{\kappa}_{j}\vec{r}-\omega_{j}t-\theta_{j})$ is a field vector potential operator which depends on $q_{j}$-expressions and a set of scaling parameters $\{\beta_{j}\}$. In the strong-field case with coherent initial state of light, when $N_{j}\gg{}1$ \blue{(and consequently $\langle{q_{j}}\rangle=\sqrt{2N_{j}}\gg{}\langle{\frac{\partial}{\partial{}q_{j}}}\rangle=0$)}, we can simply write $\vec{A}_{j}^{par}=\green{\vec{\bf \chi}_{j}}\{\beta_{j}q_{j}\}\cos(\vec{\kappa}_{j}\vec{r}-\omega_{j}t-\theta_{j})$ noting that this is nothing else than the parametric description of classical vector potential ($\{\beta_{j}q_{j}\}$ corresponding to the mode amplitudes in the classical description). \blue{This description however connects naturally both subsystems: the interacting electron subsystem $F$ with field parameters $\{\beta_{j}q_{j}\}$, and the "dressed" light field which depends on $F$ via the avereging procedure of $\hat{H}_{B}=\left\langle F\left|\hat{H}\right|F \right\rangle$.} \blue{The error of our first-order approximation in this case is proportional to some combination of $\beta_{j}^{1}$}.

Now, let us apply our approach to some simple model cases of light-electron interaction. {We consider the strong-field case, when $N_{j}\gg{}1$}. For simplicity, we consider also the so-called dipole approximation when the dependence of the vector potential on the coordinate is neglected, since $|\vec{\kappa}_{j}(\vec{r}-\langle\vec{r}\rangle_{0})|\ll{}1$, which is common for the interaction in the optical wavelength range.

First, let us consider a non-bound electron wave packet ($U(\vec{r},t)\equiv{}0$), localized in some region of space with initial momentum 0. Then, the system of Eqs.(\ref{pf13H}a-\ref{pf13H}b) can be solved analytically:
\begin{subequations}\label{pf14}
\begin{align}
\Phi(\vec{r},\{\beta_{j}q_{j}\},t)=&\exp\bigg[-\frac{1}{2}it\hat{p}^{2}+\frac{i}{c}\hat{p}\int\limits^{t}\vec{A}_{P}(\tau)d\tau\\
&-\frac{1}{2c^{2}}i\int\limits^{t}\vec{A}^{2}_{P}(\tau)d\tau\bigg]\,\Phi(\vec{r},0)\;,\nonumber\\
i\dot{G}=&\frac{1}{2}\left[-\frac{1}{c}\sum\limits_{j}\hat{\vec{A}}_{j}\right]^{2}G\;.
\end{align}
\end{subequations}
Eq.(\ref{pf14}a) describes the dynamics of the electronic wave packet in a given field vector potential $\vec{A}_{P}=\sum\limits_{j}\vec{A}_{j}^{par}(\{\beta_{j}q_{j}\},t)$ while Eq.(\ref{pf14}b) gives the correction to the initial field state $G(\{q_{j}\},0)$ due to the interaction with the electron. The solution of Eq.(\ref{pf14}b) includes terms $\exp(d_{ij}(t)q_{i}q_{j})$ \blue{(where $d_{ij}(t)$ are some time-dependent functions),} thus it demonstrates entanglement between different modes during the interaction. In case of a one-mode field, the solution of Eq.(\ref{pf14}a-\ref{pf14}b) corresponds to the closed-form solution \cite{screp} with a relative error $\sim{}\beta^{2}$.

Let us now consider the more complicated case, when the electron is bound in a quadratic potential $U=u(t)\,{\bf x}^{2}$. This situation can be related to some model bound potential ($u(t)=\text{const}$) or to the problem of an electron in magnetic fields (time-independent or time-dependent). Since, in that case, the total Hamiltonian $\hat{H}_{P}$ for the interacting system is quadratic, the solution for $\Phi$ can be obtained analytically. \blue{The analysis of the solution is beyond the scope of this article, however the main exponential part of the solution is evident:}
\begin{subequations}\label{pf15quadr}
\begin{align}
G&(\{q_{j}\},t)=\nonumber\\
&C(t)\cdot\exp\left[\sum\limits_{j}a_{j}(t)q_{j}^{2}+\sum\limits_{j}b_{j}(t)q_{j}+\sum\limits_{i\neq{}j}d_{ij}(t)q_{i}q_{j}\right]\;,
\end{align}
\end{subequations}
where the coefficients $a_{j}(t)$, $b_{j}(t)$, $d_{i\,j}(t)$ depend on the initial states of the modes, initial state of the electron, and $u(t)$. We would like to emphasize that the parameters of the exponential part Eq.(\ref{pf15quadr}) can be controlled by external parameters in various ways. This means that depending on the conditions and the binding potential, one can generate nonclassical states of light. In particular, when $(\,a_{j}<-\frac{1}{2}\,)$, it is possible to observe amplitude-squeezed states. It is possible to obtain also multi-mode squeezed states of different types. 

\blue{Finally, we would like to note our approximation in the context of multi-electron systems. The generalization is straitforward:}
\begin{subequations}\label{pf13Hn}
\begin{align}
i\dot{F}=&\left[\sum_{e}\hat{H}_{P}+\sum_{e\neq{}e'}U_{e{}e'}\right]F\;,\;\;\;\;\\
&\hat{H}_{P}=\sum_{e}\frac{1}{2m_{e}}\left[\,\hat{\vec{p}}_{e}-\frac{1}{c}\sum\limits_{j}\vec{A}_{j}^{par}\right]^{2}+\sum_{e}U(\vec{r},t)\;,\nonumber\\
i\dot{G}=&\hat{H}_{B}G\;,\;\;\;\;\hat{H}_{B}=\left\langle F\left|\hat{H}_{P}\right|F \right\rangle\;,
\end{align}
\end{subequations}
\blue{This system can be further used to unify various multi-electron theories (for example band structure theory) with quantized light description for the strong-field interaction problem. The noticeable difference with the one-electron consideration is that the fundamental scaling parameter here is $\beta\sim\left[N_{e}\cdot{}W_{e}/W_{field}\right]^\frac{1}{2}$, where $N_{e}\cdot{}W_{e}$ is the total oscillating energy of charged particles. Although, in many relevant cases it still can be very small, in some other cases, higher order approximations may be necessary.} 

--------------------------------
\\
\textit{Conclusions.}-- \Nblue{We present an approach which opens the possibilities to simplify and further study some quantum phenomena in interaction of light with complex quantum systems.} This approach is analogous to the BO-approximation for the multi-system interactions of different nature, including possible non-local interactions. The general scheme can be further improved by additional subsequent steps {beyond} the considered first-order approximation. Our approach can be applied to the general light-matter interaction problem and the various related particular problem. We obtain a system of equations connected parametrically which allows to simplify the problem significantly. We demonstrate multi-mode and light-electron entanglement during the interaction, and the possibility to generate nonclassical states of light. Our approach can explain recent experiments with intense interacting light \cite{wP0,quantExp} and shed light to some related phenomena towards strong-field quantum optics and beyond.\\

\begin{acknowledgments}
I. G. and S.G. highly acknowledge funding by the ERC consolidator grant QUEM-CHEM (772676).
\end{acknowledgments}

\section*{Appendix: COHERENT STATE DETAILS}
We consider the following function:\\
\begin{subequations}\label{i01}
\begin{align}
\Psi_{c}(q,t)=&C(t)\cdot{}\exp\left[-\frac{1}{2}q^{2}+q\,e^{-i\omega{t}+i\Theta}\sqrt{2N_{0}}\right]\;,\\
C(t)=&\frac{1}{\sqrt[4]{\pi}}\exp\left[-\frac{1}{2}N_{0}\right]\cdot\exp\left[-\frac{i\omega{t}}{2}-\frac{1}{2}N_{0}\,e^{-2i\omega{t}+2i\Theta}\right]\;,
\end{align}
\end{subequations}
where $q$ is a parameter of the light wave function (light state internal "coordinate"); $\Theta$ is an arbitrary given real number (phase of the light wave, see later); $N_{0}$ is an arbitrary given constant (average photon number, see later); $\omega$ is an arbitrary given frequency of the light mode. This function is called {{quantum coherent state of the plane wave light mode in coordinate representation}} because of its following properties: \\
1) It is an exact solution of the following Schr\"odinger equation:
\begin{equation}\label{i02}
i\partial_{t}\Psi_{c}(q,t)=\omega\hat{N}\,\Psi_{c}(q,t)=\frac{\omega}{2}\left(q^{2}-\frac{\partial^{2}}{\partial{}q^{2}}\right)\Psi_{c}(q,t)\;,
\end{equation}
which describes the time-dependent states of a light plane wave with frequency $\omega$ propagating in vacuum. {{Note, this is a Sch\"odinger representation of Schr\"odinger equation.}}\\
2) It is normalized: $\int\limits_{-\infty}^{+\infty}|\Psi_{c}(q,t)|^{2}dq\equiv{1}$.\\
3) Within the following definitions of the field operators:
\begin{subequations}\label{i03}
\begin{align}
&\hat{\vec{A}}={\bf{{x}_{0}}}\,\frac{1}{\sqrt{2}}{\beta}\,\Big[\hat{a}\,e^{i\kappa{z}}+\hat{a}^{+}\,e^{-i\kappa{z}}\Big]\;,\\
&\hat{a}=\frac{1}{\sqrt{2}}\Big(q+\frac{\partial}{\partial{q}}\Big)\;,\\
&\hat{a}^{+}=\frac{1}{\sqrt{2}}\Big(q-\frac{\partial}{\partial{q}}\Big)\;,\\
&\hat{N}=\Big(\hat{a}^{+}\hat{a}+\frac{1}{2}\Big)=\frac{1}{2}\,\Big(q^{2}-\frac{\partial^{2}}{\partial{q}^{2}}\Big)\;,\\
&\beta=c\sqrt{2\pi/\omega{V}}\,,
\end{align}
\end{subequations}
We can obtain the following quantum-mechanically averaged values:
\begin{subequations}\label{i03}
\begin{align}
&\langle{q}\rangle_{c}=\sqrt{2N_{0}}\,\cos\left(\omega{t}-\Theta\right)\;,\\
&\langle\frac{\partial}{\partial{q}}\rangle_{c}=-i\sqrt{2N_{0}}\,\sin\left(\omega{t}-\Theta\right)\;,\\
&\langle\hat{\vec{A}}\rangle_{c}={\bf{{x}_{0}}}\,\beta\sqrt{2N_{0}}\,\cos\left(\omega{t}-\Theta-\kappa{z}\right)\nonumber\\
&\;\;\;\;\;\;\;\;={\bf{{x}_{0}}}\,A_{0}\,\cos\left(\omega{t}-\Theta-\kappa{z}\right)\;,\\
&\langle\hat{N}\rangle_{c}=\frac{1}{2}+N_{0}\;,
\end{align}
\end{subequations}
where $\langle\hat{R}\rangle_{c}=\int\limits_{-\infty}^{+\infty}\Psi_{c}^{*}(q,t)\,\hat{R}\,\Psi_{c}(q,t)\,dq$.\\
\Nblue{Finally, we note that $\;\hat{a}\;\Psi_{c}=e^{-i\omega{t}+i\Theta}\sqrt{N_{0}}\;\Psi_{c}$ establishing a connection with the definition of the canonical coherent state in Sch\"odinger representation\cite{quant1}.}

\section*{Data Availability Statement}
The data that supports the findings of this study are available within the article.
\\

\end{document}